\documentclass{sig-alternate}
\usepackage{graphicx}
\usepackage{subfigure}
\usepackage{algorithmic}
\usepackage{algorithm}
\usepackage{ amsmath}
\usepackage{ amsmath}

 \newcommand{\remove}[1]{}

\numberofauthors{2}
\newcommand{\KL}[1]{\textbf{/*KL: #1*/}}

\begin{document}
%
\title{Impact of Dynamic Interactions on Multi-Scale Analysis of Community Structure in Networks}

\author{
\alignauthor Rumi Ghosh\\
       \affaddr{Computer Science Department}\\
       \affaddr{University of Southern California}\\
       \affaddr{Los Angeles, CA }\\
       \email{rumig@usc.edu}
\and  
\alignauthor
Kristina Lerman\\
       \affaddr{Information Sciences Institute}\\
       \affaddr{University of Southern California}\\
       \affaddr{Marina del Rey, CA 90292}\\
       \email{lerman@isi.edu}
}

\maketitle
\begin{abstract}
To find interesting structure in networks, community detection algorithms have to take into account not only the network topology, but also dynamics of interactions between nodes. We investigate this claim using the paradigm of synchronization in a network of coupled oscillators. As the network evolves to a global steady state, nodes belonging to the same community synchronize faster than nodes belonging to different communities. Traditionally, nodes in network synchronization models are coupled via one-to-one, or conservative interactions.
However, social interactions are often one-to-many, as for example, in social media, where users broadcast messages to all their followers. We formulate a novel model of synchronization in a network of coupled oscillators in which the oscillators are coupled via one-to-many, or non-conservative interactions. We study the dynamics of different interaction models and contrast their spectral properties.
To find multi-scale community structure in a network of interacting nodes, we define a similarity function that measures the degree to which nodes are synchronized and use it to hierarchically cluster nodes. We study real-world social networks, including networks of two social media providers. To evaluate the quality of the discovered communities in a social media network we propose a community quality metric based on user activity. We find that conservative and non-conservative interaction models lead to dramatically different views of community structure even within the same network.
Our work offers a novel mathematical framework for exploring the relationship between network structure, topology and dynamics.
\end{abstract}

\section{Introduction}
{Modular structure} is an important characteristic of complex real-world networks, including social networks which are composed of communities and sub-communities of interconnected individuals, and biological networks, which are often organized within functional modules~\cite{Ravasz2002Hierarchical,Rives03}. Conductance minimization~\cite{Chung1997Spectral} and modularity maximization~\cite{Fortunato2010Community} are some of the most popular methods for community detection. However, these are combinatorial approaches that have been shown to be NP-hard or NP-complete. As the result, researchers resort to heuristics and approximation algorithms when applying these methods to community detection problem.
On the other hand, decentralized algorithms based on local computation have been shown to provide scalable solutions to combinatorial problems~\cite{Yokoo98}.
Motivated by this idea, we cast community detection as a decentralized computation problem in which a network of \emph{locally} interacting agents over time finds a \emph{global} solution that corresponds to the community division of the network.


Network's modular  structure is a product of both the {topology} of its underlying connections and also its {function}, which is determined by the dynamic processes taking place on the network.
Nodes are not static but change their state or activity levels in response to the actions of neighbors.
Existing community detection algorithms focus solely on network topology and ignore dynamic processes taking place on the network, or they implicitly assume that interactions between nodes are mediated by a conservative process similar to heat diffusion~\cite{Chung1997Spectral, kondor2002}.
However, this assumption may not be justified for social networks~\cite{Ghosh10snakdd}.


In this paper we study a framework for multi-scale analysis of community structure in networks that explicitly takes interactions into account.
We consider a static network of active nodes (or agents), who can affect the state or activity of their neighbors through interactions. 
We differentiate between one-to-one interactions, henceforth referred to as \emph{conservative}, and one-to-many, or \emph{non-con\-ser\-va\-tive} interactions. Examples of the former include money exchange, Web surfing, and diffusion in physical systems. Non-con\-ser\-va\-tive interactions include broadcast-based interactions that lead to information diffusion, epidemics, and other social phenomena.
These local interactions cause nodes' activity to become more similar. In a social network, for example, frequent contact leads to similarity of behavior among friends.  Over time, communities composed of individuals who act in a similar manner will emerge. As another example, consider a population of fireflies who have characteristic light flashing patterns to help males and females recognize each other. Some firefly species exhibit synchronous flashing, during which individual's flashing pattern can affect that of his neighbors, leading all nearby fireflies to flash in unison~\cite{Strogatz-book}. The Kuramoto model is a simple mathematical description of distributed synchronization in this and other physical and biological systems~\cite{Kuramoto}. The model considers a network of coupled oscillators, in which the phase of each oscillator is affected by the phases of its neighbors. While the network as a whole eventually reaches a fully synchronized state, it does so in stages, with nodes belonging to the same community synchronizing faster than nodes belonging to different communities~\cite{Arenas06}.

We propose a new model of distributed synchronization based on non-conservative interactions. We show that in this {interaction model}, nodes synchronize much faster than in the conservative interaction model.
We use dynamic interaction models to explore community structure of several networks, including a benchmark social network and  large real-world networks from social media sites.
We investigate how the dynamics of synchronization, and  the community structure that emerges from it, are affected by the nature of interactions.
Our study reveals substantial differences in network structure discovered by different {interaction} models. We find a complex layered organization of the real-world networks.
While these networks exhibits the `core and whiskers' organization found in other real-world social and information networks~\cite{Leskovec08www}, with a giant core and multiple small communities (whiskers) weakly connected to the core, this constitutes but one layer of the organization. As we peel away the whiskers layer to examine the core, we find a similar `core and whiskers' structure in the new layer, and so on.  We evaluate community division by measuring similarity of community members, introducing an activity-based measure of similarity for social media users.
We show that non-conservative interaction model finds many more valid communities.

The principal contribution of this paper is a formal framework that consolidates and generalizes approaches to community detection in complex networks.  We apply  this framework to study real-world networks.
The specific contributions that support this perspective are
\begin{itemize}
\item A novel model for locally interacting nodes coupled via non-conservative interactions
\item A methodology for hierarchical community detection based on synchronization similarity and an activity-based measure of community quality
\item Detailed investigation of interaction models on real-world networks revealing important differences between them within a layered `onion-like' organization of complex networks
\end{itemize}

While it may seem counter-intuitive that a network's community structure depends on anything but its topology, as we show in this paper, different dynamic processes running on the same topology can lead to different views of network structure. In reality network structure, its topology and dynamics are intricately interconnected and our work offers a formal framework to begin exploring these connections.



\section{Network Interaction Models}
\label{sec:synchronization}
We consider a network of active nodes (e.g., agents or actors), each interacting locally with its neighbors.
Interactions between nodes determine the dynamic process taking place on the network. Consider financial exchange networks in which nodes distribute money among their network neighbors. The interactions that give rise to the financial exchange can be called \emph{conservative}, since they do not increase nor decrease the amount of money exchanged.
Web surfing, communicating via phone calls, and other one-to-one interactions are conservative, because, as in the case of a Web surfer, at any time the surfer can browse only one page, and the probability to find the surfer on any Web page remains constant.
We contrast these to \emph{non-conservative} interactions, which do not preserve the amount of quantity exchanged. Take, as an example, a virus spreading through a social network. A person (node) will get infected with a virus through her infected friends, but the amount of the virus present in the network will increase because of these interactions (or decrease as infected people become cured). Social processes based on one-to-many interactions, such as users broadcasting messages in online social media, are also non-conservative in nature. While the conservative/non-conservative dichotomy might not capture the full range of possible interactions in a network, we begin our investigation here because this dichotomy can be described mathematically. Moreover, to keep mathematics tractable, we focus analysis on linear interactions.

Physicists have studied the dynamics of interacting entities in an attempt to understand collective behavior of complex networks. The Kuramoto model~\cite{Kuramoto} was proposed as a simple model for how global synchronization may arise in physical and biological systems. The model considers a network of phase oscillators, each coupled to its neighbors through the sine of their phase differences. The Kuramoto model has a fully synchronized steady state in which the phase difference between all oscillators is zero.

As we show below, the Kuramoto model (at least in the linear case) assumes that interactions between nodes are mediated by a conservative process similar to heat diffusion, which is mathematically related to the random walk. However, not all social phenomena, including epidemic spread and information diffusion, admit to such descriptions~\cite{Ghosh10snakdd}. In this section we introduce a new model of distributed synchronization based on non-conservative interactions.

\subsection{Conservative Interaction Models}
The Kuramoto model is written as:
\begin{equation}
\frac{d\theta_i}{dt} =\omega_i+\sum_{j \in neigh(i)} K_{ij}  sin(\theta_j-\theta_i)  \label{eq:31}
\end{equation}
\noindent where $\theta_i$ is the instantaneous  phase of the $i$th oscillator, $\omega_i$ is its natural frequency, and $K_{ij}$ is the coupling constant that describes the strength of interaction with $j$th neighbor. The neighborhood of node $i$, $neigh(i)$, contains nodes which share an edge with node $i$.
For small phase differences, $sin \theta \approx \theta$, and the linearized version of the Kuramoto model can be written as:
\begin{equation}
\frac{d\theta_i}{dt} =\omega_i+\sum_{j \in neigh(i)} K_{ij}  (\theta_j-\theta_i)  \label{eq:3}
\end{equation}

In a more general sense, we treat $\theta_i$ as some extrinsic property of node $i$ (agent), which is dynamic and can be affected by the local interaction with the neighbors. The quantity $\omega_i$ can then be perceived as its intrinsic property, which is not affected by external factors and remains constant over time. For example, $\theta_i$ could represent the opinions of an individual agent $i$, and $\omega_i$ his intrinsic beliefs. Though his opinions depend on his intrinsic beliefs, they may change over time as the result of interactions with neighbors. Though rather simplified, we believe that this abstract model provides a useful framework to study social phenomena.

For convenience, we rewrite Eq.~\ref{eq:3} in vector form:
\begin{equation}
\label{eq:1}
\frac{d\theta}{dt}=\omega - K\cdot L\theta
\end{equation}
\noindent Here $\omega$ is the vector of length $N$ of intrinsic properties of nodes, $\theta$ is a vector of  their extrinsic properties, and $K$ is a matrix of pairwise couplings constants between nodes. $K\cdot L$ is the dot product of $K$ and $L$.
Operator $L$ is the Laplacian of the graph $L=D-A$. Here $A$ is the adjacency matrix of the unweighted, undirected graph, such that $A[i,j]=1$ if there exists an edge between $i$ and $j$; otherwise, $A[i,j]=0$. Matrix $D$ is the diagonal matrix where $D[i,i]=\sum_i A[i,j]$ and $D[i,j]=0$ $\forall$ $i\neq j$.

The model describes evolution of the extrinsic properties  of a population of nodes (or agents). After some time, the network reaches a steady state, and interactions no longer change the property of any node, i.e., ${\theta}_i(t)={\theta}_{i}(t+1)$. In the opinion formation example, it would mean that after some period, individual opinions no longer change. For $\omega_i=\omega_j,\ \forall i,j$,  in the steady state $\theta_i(t)=\theta_j(t),\ \forall i,j$. In other words, the extrinsic properties of  all the nodes are the same in the steady state. In the context of oscillators, this means that their phases are equal and they are synchronized.

To see why the linearized Kuramoto model is conservative, we imagine that interactions result in node exchanging some content with neighbors.
Imagine that at time $t$, node $i$ has an amount $\theta_i(t)$ of content and produces some amount $\omega_i$ for itself and some amount $d_i\theta_{i}(t)$ for its neighbors, which it then transfers to its $d_i$ neighbors (transmission is denoted by negative sign in Eq. \ref{eq:3}). Each neighbor receives $1/{d_i}$ of the transmitted amount (reception is denoted by positive sign in Eq. \ref{eq:3}). Thus, whatever is produced is completely transferred to other nodes in the system. 



The Kuramoto model is just one of a family of conservative interaction models.
The model would change based on the nature of interactions. In the case when the new amount of content produced by node $i$  at each time step is $\theta_i$ (instead of  $d_i\theta_i$ in Eq. \ref{eq:1} ), the conservative interaction model becomes:
\begin{equation}
 \frac{d\theta}{dt}=\omega-K\cdot(I-AD^{-1})\theta
 \label{eq:61}
 \end{equation}
 Here, $D^{-1}$ is the inverse of the diagonal matrix.
 Another conservative interaction model could be framed using the normalized Laplacian  operator:
 \begin{equation}
 \frac{d\theta}{dt}=\omega-K\cdot(I-D^{-1/2}AD^{-1/2})\theta
 \label{eq:62}
 \end{equation}
The normalized Laplacian operators in Eq. \ref{eq:61} and \ref{eq:62} is often used to describe random walk-based processes.
Eq.~\ref{eq:1} has been used to describe a variety of conservative systems. When $\omega=0$ and $K[i,j]= c,\ \forall i,j$, it measures electric potential in a network of capacitors  of unit capacitances, with one plate of each capacitor grounded and the other plate connected according to the graph structure, with each edge corresponding to a  resistor of resistance $\frac{1}{c}$.
The same equation (with $\omega=0$, $K[i,j]=c$) has been used to model (discrete) diffusion of heat and fluid flow in networks and serves as the basis of diffusion kernels over discrete structures in machine learning algorithms \cite{kondor2002}.

\subsection{Non-conservative Interaction Models}
In contrast to conservative interaction models, in most human or biological networks what is produced is not necessarily completely transferred or distributed among neighbors. Some portion of it might be dissipated or lost. This changes the nature of interactions and the resulting dynamics of the network.
We present a model of non-conservative interactions in undirected networks:
\begin{eqnarray}
\frac{d\theta_i}{dt} &=&\omega_i+\sum_{j \in neigh(i)} K_{ij}  (\theta_j-\frac{\alpha \theta_i}{d_i})\label{eq:4} \\
\frac{d\theta}{dt}&=&\omega -K\cdot(\alpha I-A)\theta
\label{eq:2}
\end{eqnarray}
Here $\alpha$ is a constant and  $I$ is the identity matrix. The equation above introduces a new operator, which we call the \emph{Replicator operator} $R=\alpha I-A$.   In order for this system to reach a steady state, $\alpha  \ge \lambda_{max}$ where  $\lambda_{max}$ is the largest eigenvalue of the adjacency matrix of the network. Again without loss of generality we can take $K_{ij}=c$. Eq.~\ref{eq:2} gives the vector form of the non-conservative model.

As in the conservative interaction model, we can imagine that at each time step, node $i$ produces some amount  $\omega_i$ of content for itself. In addition, it produces $\alpha \theta_i$ of additional content and transmits it to the system regardless of the actual number of neighbors it has (transmission is denoted by negative sign in Eq.~\ref{eq:4}). Each neighbor receives an amount $\theta_{i}$  from the system (reception is denoted by positive sign in Eq.~\ref{eq:4}). Thus $(\alpha-d_i)\theta_i$ of the new content created by node $i$ is not transferred to any neighbor and is lost. This accounts for non-conservation during interactions. In spite of non-conservation, the system reaches a steady state where phases of oscillators no longer change: ${\theta}_i(t)={\theta}_i({t+1})$. In steady state, $\theta_i$ is proportional to the $i^{th}$ element of the largest eigenvector of the adjacency matrix.



Other flavors of the non-conservative interaction model are also possible. If the amount produced by node $i$  at each time step is $\theta_t[i]$ (instead of  $\alpha \theta_{i}$ in Eq. \ref{eq:2} ), another non-conservative  linear interaction model  could be:
  \begin{equation}
 \frac{d\theta}{dt}=\omega-K\cdot(I-\alpha^{-1}A)\theta
 \label{eq:7}
\end{equation}
If $\alpha  \ge \lambda_{max}$, then this system would reach equilibrium.

\subsection{Spectral Properties of Operators}
As we saw above, the linear conservative model naturally gives rise to the Laplacian operator $L$ (see Eq.~\ref{eq:1}). This explains the connection between the spectrum of the Laplacian and topological properties of synchronized structures that emerge as the network evolves to the fully synchronized state. The number of null eigenvalues of $L$ gives the number of disconnected components of the graph and is the basis of spectral clustering. The time to reach the steady state is inversely proportional to the smallest positive eigenvalue of the Laplacian, and the gaps between consecutive eigenvalues are related to the relative difference in synchronization time scales of different modules~\cite{Arenas06,Arenas2008Synchronization}.

The replicator operator $R$ we introduced in Eq.~\ref{eq:2} is the non-conservative counterpart of the Laplacian. Its spectrum gives us information about topological and temporal scales of non-conservative dynamical systems. In particular, the time it takes for the system to reach the steady state is inversely proportional to the smallest positive eigenvalue of $R$
(when $\alpha=\lambda_{max}$)
as shown in the next section.

\subsection{Generalized Interaction Model}
\label{gim}
Both conservative and non-conservative interaction models are special cases of the general linearized interaction model. As we show below, this model generalizes several community detection methods, such as spectral clustering, modularity maximization and conductance minimization.

A generalized linear model of interaction can be written in terms of the {operator} $\mathcal{L}(A)$ of the adjacency matrix $A$.
\begin{equation}
 \frac{d\theta}{dt}=\omega-K\cdot \mathcal{L}(A)\theta
 \label{eq:5}
 \end{equation}
Solving this differential equation we get:
\begin{equation}
\theta({t})=(\theta_0-{(K\cdot \mathcal{L}(A))}^{-1}\omega) e^{-K\cdot\mathcal{L}(A)t} +{(K\cdot\mathcal{L}(A))}^{-1}\omega
 \label{eq:8}
 \end{equation}
\noindent with $\theta_0$ the initial value of $\theta(t=0)$, and $\omega$ the vector of natural frequencies.

Let $|V|$ be the number of nodes in the network.  Let $\mathcal{X}$ be a $|V|\times|V|$ matrix
whose column $\mathcal{X}[.,i]$ gives the eigenvector of $\mathcal{L}(A)$ corresponding to eigenvalue $\lambda_{i}$. Also, let $\Lambda$ be the diagonal eigenvalue matrix where $\Lambda[i,i]=\lambda_{i}$. Let $\mathcal{Y}={\mathcal{X}}^{-1}$. Therefore $\mathcal{L}(A)= \sum_{i \in \{1,2,\cdots |V|\}}\mathcal{X}[.,i]\lambda_i \mathcal{Y}[i,.]$.
 Eq. \ref{eq:8}  with $\omega=0$, $K[i,j]=c$ can be rewritten as:
 \begin{eqnarray}
\theta({t})&=&\theta_{0}e^{-c{\mathcal{L}(A)}t} \nonumber \\
&=& \sum_{i \in \{1,2,\cdots |V|\}}\mathcal{X}[.,i]e^{-c\lambda_i t}\mathcal{Y}[i,.]\theta_{0} \nonumber \\
&=&\sum_{i \in \{1,2,\cdots |V|\}}\mathcal{X}[.,i]e^{-c\lambda_i t}c_i
 \label{eq:9}
 \end{eqnarray}
\noindent Here $c_i=\mathcal{Y}[i,.]\theta_{0}$ is a constant. Let $\lambda_1 \le \lambda_2\le\cdots \lambda_{max}$.
Let  $t_j$ be  such that $e^{-c\lambda_i t_j} \to 0,\ \forall i \ge j$ and $t_{j+1}$ be such that $e^{-c\lambda_i t_{j+1}} \to 0,\ \forall i \ge j+1$. Therefore, for  $t_{j+1} \le  t < t_{j}$, $\theta_{t}=\sum_{i=1}^{j}\mathcal{X}[.,i]e^{-c\lambda_i t}c_i$.

\paragraph{Steady State}
Let us look at the $\lambda_1=0$ case more closely. This arises  in
non-conservative interaction when $\alpha=\lambda_{max}$ (Eqs.~\ref{eq:2} and \ref{eq:7}).
In this case as $t\to \infty$, Eq.~\ref{eq:9} reduces to $\theta_{t \to \infty}= \mathcal{X}[.,1]c_1$, where $c_1$ is a constant which is the steady state or equilibrium.
For non-conservative interaction models,  $\lambda_1>0$  leads to a trivial equilibrium condition.
 Considering Eq. \ref{eq:1}, \ref{eq:2}, \ref{eq:61}  and \ref{eq:7} with
 $\omega=0$:
\begin{description}
\item [$\mathcal{L}=D-A=L$:]
In this case $\mathcal{X}[.,1]= \bar{1}$ ( vector of 1s). Hence $\theta_{t \to \infty}[i]=\theta_{t \to \infty}[j]$ $\forall$ $i,j$. Hence the content or phase of all nodes is equal at synchronization.
\item [$\mathcal{L}=I-AD^{-1}$:]: Hence $\theta_{t \to \infty}[i] \propto d[i]$ where $d[i]$ is the degree of node $i$.
\item[$\mathcal{L}=I-\frac{1}{\lambda_{max}}A$ or $\mathcal{L}=\lambda_{max}I-A=R$:]  Here $\theta_{t \to \infty}$ $\propto$ the eigenvector of the adjacency matrix A corresponding to the largest eigenvalue.
\item[$\mathcal{L}=I-\frac{1}{\alpha}A $  or  $\mathcal{L}=\alpha I-A $ $\forall \alpha > \lambda_{max}$ :]  Here $\theta_{t \to \infty}[i] \to 0 \ \forall i$
\end{description}

\paragraph{Spectral Clustering and Partitioning } Note that if  $\mathcal{X}[.,i]$, $ \forall i \in \{1,\cdots, j\}$ is used for clustering, the conservative interaction models in Eq. \ref{eq:1} ,\ref{eq:61} and \ref{eq:62}, reduce to spectral clustering techniques using Laplacian (Eq. \ref{eq:1}) or normalized Laplacians (Eq. \ref{eq:61} or \ref{eq:62} )  to find $j$ communities \cite{tutorialspectral}. \cite{Spielman96spectralpartitioning} showed that naive spectral bisection methods do not necessarily work. However,  for a conservative dynamic process  with $\mathcal{L}=D-A$ (Eq. \ref{eq:1} with $\omega=0$), if vertices are arranged such that $\theta_{t}[u_1]\ge \theta_{t}[u_1]\ge \cdots \theta_{t}[u_{|V|}]$ and  set $S_i$ comprise of nodes $u_1,u_2\cdots u_{i}$, then  at  $t=t_{2}: e^{-c\lambda_i t_2} \to 0$, $\forall i > 2$, $\min_{S_i} \frac{E(S_i,\bar {S_i})}{\min(S_i \bar{S_i})}$(Fiedler cut) is $O(1/\sqrt n)$  for bounded degree planar graphs and a bisector of $O(\sqrt{n})$ can be found by repeatedly finding Fiedler cuts. (This is one of the very few theoretical guarantees for spectral partitioning.)

\paragraph{Conductance}
Finding a partition with a low conductance is closely related to the conservative interaction model with  $\mathcal{L}=I-AD^{-1}$(Eq.  \ref{eq:61} with $\omega=0$ ).
Let $S \subset V$ be a set of vertices. Let $E(S,\bar S)$  be the cut size or the edges going from $S$ to $\bar S$. Volume $vol(S)$ is the sum of the degree of all vertices in S. Conductance is given by $\phi(G)=\min_S \frac{E(S,\bar S)}{vol(S)}$. The classic Cheeger inequality states that $2\phi(G) \ge \lambda_2 \ge \frac{\phi(G)^2}{2}$. Therefore, if this conservative dynamic process  starts at node $u$, i.e., $\theta_0[u]=1$ for $u \in V$ and $\theta_0[v]=0$ $\forall v\neq u \in V$, then $|\theta_{t}[v]-\theta_{\infty}[v]|\le e^{-t\frac{\phi(G)^2}{2}} \sqrt{\frac{d[u]}{d[v]}}$ where $d[u]$ is the degree of node $u$. In other words, if  conductance is large, this dynamic process would reach equilibrium quickly.  Let the nodes be arranged such that $\frac{\theta_t[u_1]}{d[u_1]} \ge \frac{\theta_t[u_2]}{d[u_2]}\ge \cdots \ge \frac{\theta_t[u_{|V|]}}{d[u_{|V|}]}$ at time t, and let set $S_i$ comprise of nodes $u_1,u_2\cdots u_{i}$.
In this setup,  for a set with volume $vol(S)\le \frac{vol(G)}{4}$ and  $\phi(G)<\gamma$, where $\gamma$ is a constant, there is a subset $S' \subset S$ with volume $vol(S') \ge vol(S)/2$, such that, if the conservative dynamic process (Eq.  \ref{eq:61} ) starts at $u \in S'$, at $t=\lceil \frac{\gamma^{-2}}{4} \rceil$,  $\min_{S_i} \frac{E(S_i,\bar {S_i})}{vol(S_i)} \le \gamma \sqrt{log(vol(S)}$. This shows that by focussing on cuts determined by linear ordering of vertices using $\theta_t$ of  conservative interaction model in Eq. \ref{eq:61}, the partition obtained is quadratic factor of the minimum conductance (which is one of the best approximation guarantees for local partitioning using conductance)~\cite{Chung1997Spectral}.


\paragraph{Modularity Maximization} If $\mathcal{L}(A)=DD-A$ where $DD[i,j]=\frac{d[i]d[j]}{2m}$ where $d[i]$ is the degree of node $i$ and $d[j]$ is the degree of node $j$ and $2m$ are the total number of edges, and if  $\mathcal{X}[.,i], \forall i$ is used for clustering, then the model reduces to modularity maximization problem using the eigenvector approach \cite{newmaneigenvector}.


In the section below we use interaction models to identify community structure that emerges en route to the steady state in real-world networks.
We find that conservative and non-conservative interaction models lead to similar multi-scale organization of the network, but the composition of communities found at different scales is markedly different.

\section{Interaction Dynamics and Community Structure}
\label{sec:communities}
A community is a group of nodes who are more similar to each other than to other nodes. Some network community detection approaches
like conductance
 measure similarity by the number (or fraction) of edges linking nodes to other nodes within the same community~\cite{Fortunato2010Community}. The interaction models allow us to define communities dynamically. Given a network of nodes with random initial states ($\theta_i(t=0)$), we allow the system to evolve according to   the rules of the interaction model.  As Arenas et al.~\cite{Arenas06} observed, as nodes interact, their phases (or extrinsic properties) become more similar, with nodes within the same community becoming more similar to each other faster than nodes from different communities. This happens in stages that reveal the network's hierarchical community structure. In this section we define a new similarity function and describe a hierarchical clustering algorithm that uses it to identify a network's community structure.



\paragraph{Similarity Measure}
We assume that when nodes are similar, further interactions between them do not change their extrinsic property, which is given by the dynamic variable $\theta_i(t)$.
Maximal similarity is reached at time $t^{eq}$, when the equilibrium or steady state is reached. 
In the conservative model in Eq.~\ref{eq:1}, the steady state corresponds to global synchronization, in which every node has the same phase at any time if the natural frequencies of all nodes are equal, i.e.,  $\omega_i=\omega,\ \forall i$. The steady state of the non-conservative model is given by the largest eigenvector of $R$ (or the adjacency matrix $A$) when $\omega_i=0, \forall i$. For the sake of convention, we call this state the synchronized state, even if the values of all $\theta_i$s are not the same (but they do have fixed values, given by the first eigenvector). Once the system reaches synchronization, $\theta_i(t+1)=\theta_i(t)$ for all subsequent times.

Arenas et al. used cosine of the phase difference between nodes as the measure of similarity. However, such a measure will lead to finite differences between nodes in the steady state in the non-conservative model. Instead, we measure similarity by the relative difference of the variables in the synchronized state. In other words, similarity between nodes $i$ and $j$ at time $t$ is
$$sim(i,j,t)=cos(\theta_i(t) - \frac{\theta_i^{eq}}{\theta_j^{eq}}\theta_j({t})) $$
\noindent where $\theta_i^{eq}$, is the value of the dynamic variable in the steady state.
Therefore for both the conservative and non-conservative interaction models, $sim(i,j,t)=1$, $\forall$ $i,j \in V$ at $t \ge t^{eq}$.


In the conservative case, the similarity measure we propose reduces to the one used by Arenas et el., because in the conservative steady state $\theta_i^{eq}=\theta_j^{eq}$; therefore,   $sim(i,j,t)=cos(\theta_i(t) - \theta_j({t})) $.

\paragraph{Hierarchical Community Detection}
We simulate the interaction model by letting the network evolve from some initial configuration.
At any time $t<t^{eq}$, we can find the structure of the evolving network by executing  a clustering algorithm, e.g., average link hierarchical agglomerative algorithm, with the similarity calculated as shown above.
\remove
{
Algorithm \ref{alg:1} gives the hierarchical clustering algorithm.
\KL{This is a generic average link hierarchical clustering algorithm. Let's cite it and save space.}
\begin{algorithm}
\caption{Hierarchical structure of an evolving network}
\label{alg:1}
\begin{algorithmic}
\STATE{\bf{Input}}\\
$K$: number of simulations of the interaction model $\mathcal{I}$ \\
$t$: time at which the hierarchy is calculated\\
$\theta_k(t)$: $\theta(t)$ from the $k^{th}$ simulation  \\
$\bar \theta(t)= (\theta_i(t)[1], \theta_i(t)[2], \cdots, \theta_i(t)[K])$.\\
\STATE{\bf{Output}}\\
The hierarchical structure of the community using interaction model $\mathcal{I}$.
\STATE  {\bf{Initialize }}\\
Assign each node $i$ to a separate community $C_{i} \in C$.\\
\REPEAT
\STATE Find the centroid of each community $C_i$, $cen(C_{i}(t))$ \\
  \STATE Find communities, $C_i$ and $C_j$ such that $sim(cen(C_{i}(t), cen(C_{j}(t), t) = \frac{1}{K}{\sum_{i=1}}^{K}cos(cen(C_{i}(t)) - \frac{cen(C_{i}(t^{eq}))}{cen(C_{i}(t^{eq}))}cen(C_{j}(t))) $ is maximum. \\
  Merge $C_i$ and $C_j$\\
   \UNTIL{$|C|=1$}
\end{algorithmic}
\end{algorithm}
}

The hierarchical structure of the network can be captured by a dendrogram. However, a complete dendrogram may be difficult to visualize, especially for large networks. Instead, we use a coarse-graining strategy to cluster nodes if their similarity is above some threshold $\mu$.
Algorithm~\ref{alg:2} describes the clustering procedure that takes similarity threshold $\mu$ as input, and at time $t$ finds all communities in the network, such that if $i \in C_i$, $\max_{j \in C_i}(sim(i,j,t))$  is more than or equal to   $1-\mu$. Since by construction, in Algorithm \ref{alg:2},  for every $i \in C_i$, there exists a $j \in C_i$ , $1-\mu \le sim(i,j,t) \le \max_{j \in C_i}(sim(i,j,t))$, therefore in all communities output by this algorithm, for  all nodes $i \in C_i$, similarity $\max_{j \in C_i}(sim(i,j,t))\ge(1- \mu)$. This algorithm has linear runtime,  $O(|E|)$, where $|E|$ is the number of edges. By changing $\mu$, we can change the number and size of clusters. As $\mu$ increases, a cluster fragments into sub-clusters and thus a hierarchical arrangement of the clusters can be found.
The set of communities  output by Algorithm \ref{alg:2} at time $t$, for a given $\mu$ is unique  and independent of the order in which edges $e(i,j) \in E$ are considered (proof omitted due to space constraints). \remove{Appendix Theorem \ref{th1}} 

\begin{algorithm}
\caption{ Communities at time $t$ with threshold of similarity, $\mu$}
\label{alg:2}
\begin{algorithmic}
\STATE{\bf{Input}}\\
$K$: number of simulations of the interaction model $\mathcal{I}$ \\
$t$: time at which the hierarchy of the evolving communities is calculated\\
$\theta_i(t)[k]$: $\theta_i(t)$ from the $k^{th}$ simulation \\
$\bar \theta_i(t)= (\theta_i(t)[1], \theta_i(t)[2], \cdots, \theta_i(t)[K])$.\\
$\mu=$similarity threshold \\
$G(V,E)=$ network with $|V|$ nodes, $|E|$ edges \\
$e(i,j)$=edge between $i$ and $j$ \\
\STATE{\bf{Output}}\\
Communities \{$C_i$\} such that $\forall i \in V$ $\max_{j \in C_i}(sim(i,j,t)) \ge (1-\mu)$ in the interaction model $\mathcal{I}$.\\
\STATE  {\bf{Initialize }}\\
$S=E$\\
Assign each node $i$ to a separate community $C_i \in C$.\\
\REPEAT
\FOR {each $e(i,j) \in E$}
    \STATE  $sim(i,j,t)= \frac{1}{K}\sum_{y=1}^{K}cos\big(\theta_i({t})^{[y]} - \big(\frac{\theta_{i}^{eq}}{\theta_{j}^{eq}}\big)^{[y]}\theta_j({t})^{[y]}\big) $
    \STATE $S=S-\{e(i,j)\}$
    \IF {$sim(i,j,t) \ge (1-\mu)$}
       \STATE Merge $C_i$ and $C_j$
    \ENDIF
\ENDFOR
\UNTIL{$S=\phi $}
\end{algorithmic}
\end{algorithm}
\paragraph{Fast and scalable}
The decentralized nature of the interaction models allows each node $i$ to compute $\theta_i$ locally interacting with at most $d[i]$ of its neighbors, which helps us to parallelize the computation process making it fast and scalable.   Due to the linear nature of the interaction models considered, Eq. \ref {eq:5} can easily be rewritten as  $\sum_{i=1}^{|V|}\frac{d \theta(i)}{dt}=\omega-K\cdot \mathcal{L}(A) \theta(i) |_{\bar \theta_0(i)}$ where $\bar \theta_0(i)[i]=\theta_0[i]$ and is  $\bar \theta_0(i)[j]=0$ $\forall$ $j\neq i$, $\theta_0$ being the initial starting vector in Eq. \ref {eq:8}. Each of the $|V|$ terms of this model can be calculated independently increasing  parallelizability further.

\section{Empirical Study}
\label{sec:validation}
We study the structure of real-world networks including Digg and Facebook. We contrast the structure discovered by the linearized Kuramoto model, given by Eq.~\ref{eq:1}, to that discovered by the non-conservative interaction model, given by Eq.~\ref{eq:2}. In each simulation, the initial phases of nodes are drawn from a uniform random distribution $[-\pi,\pi]$ and all $\omega$s are set to 0. In the non-conservative interaction model we took $\alpha=\lambda_{max}$
i.e. interaction models which reach non-trivial equilibrium ($\theta_{t \to \infty}[i] \propto$ the eigenvector of the adjacency matrix $A$ corresponding $\lambda_{max}$). Investigations into the differences of  interaction models reaching trivial equilibrium  $\theta_{t \to \infty}[i] =0$ $\forall i$ and those reaching non-trivial equilibrium is the scope of future work.
We ran multiple simulations ($O(100)$) of each interaction model with different initial conditions and use these as input to the structure detection algorithms described in the previous section.

\subsection{Karate Club}
\begin{figure*}[htp]
\begin{tabular}{@{}c@{}c@{}c@{}c@{}}
\includegraphics[height=0.24\textwidth,width=0.24\textwidth]{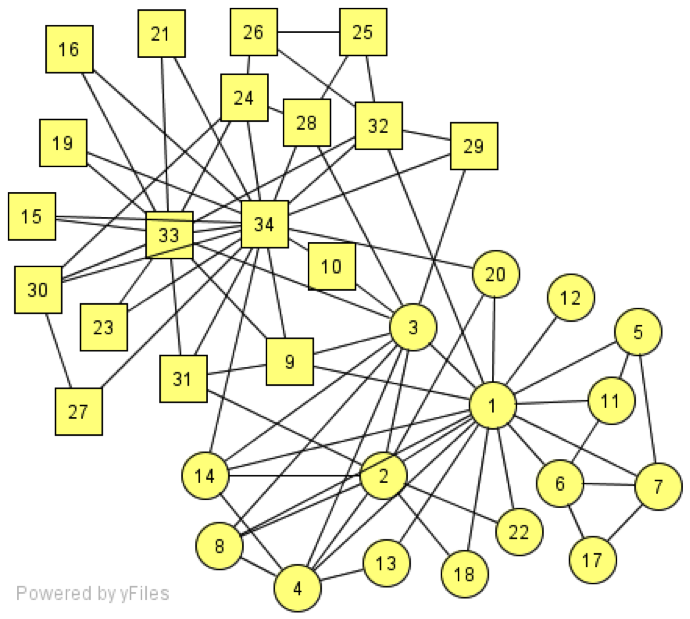}&
\includegraphics[height=0.26\textwidth,width=0.26\textwidth] {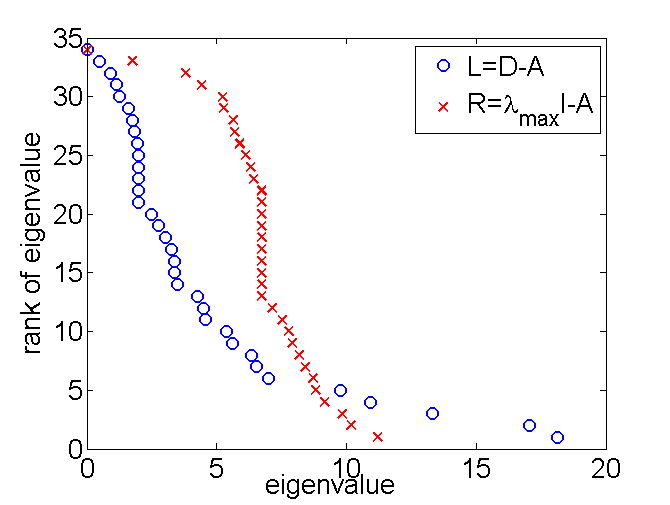} &
\includegraphics[height=0.26\textwidth,width=0.26\textwidth]{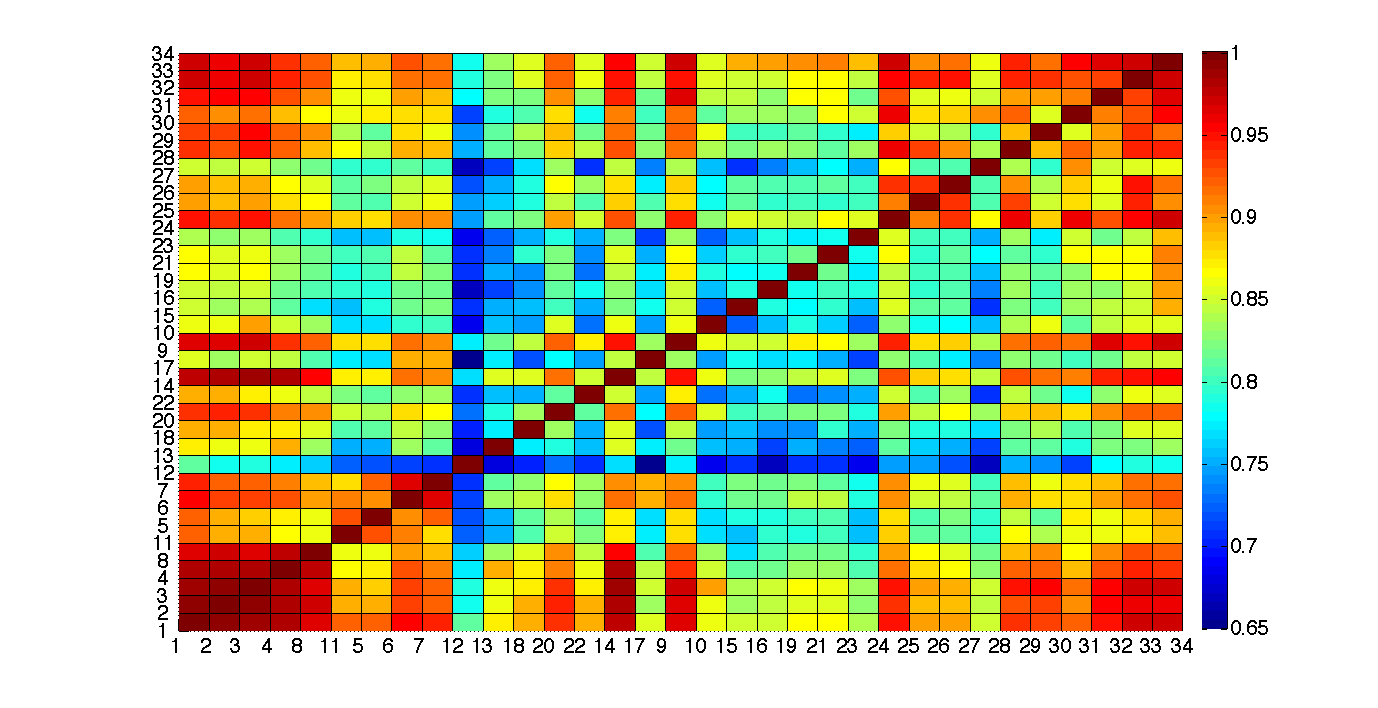} &
\includegraphics[height=0.26\textwidth,width=0.26\textwidth]{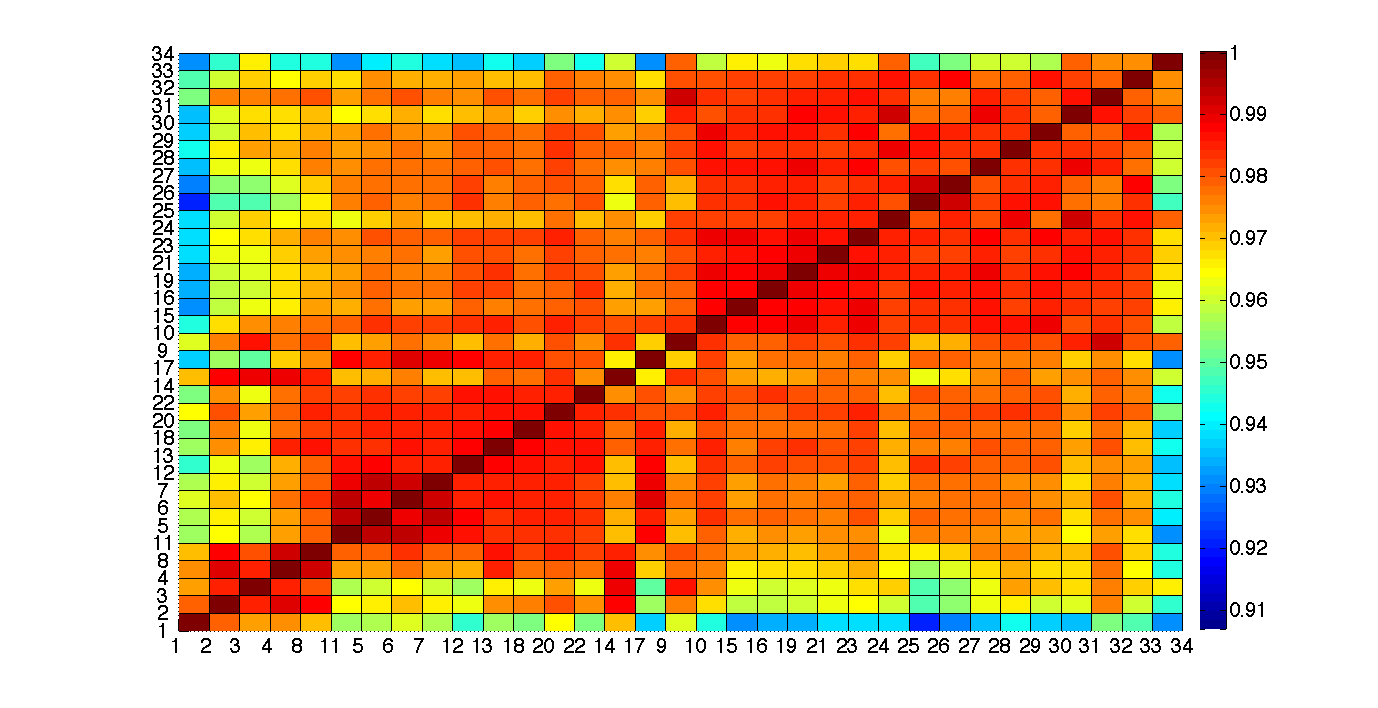} 
\\
(a)  & (b)   & (c) &(d)
\end{tabular}
\caption{Analysis of the karate club network. (a) Friendship graph. (b) Comparison of eigenvalues of the Laplacian and Replicator operators. Synchronization matrix at time $t=1000$ due to (c) the conservative interaction model and (d) the non-conservative interaction model. The color of each square indicates how similar two nodes are (zoom in to see node labels), with red corresponding to more similar nodes and blue to less similar nodes.
}
 \label{fig:karate}
\end{figure*}

We study the real-world friendship network of Zachary's karate club~\cite{Zachary},  shown in Fig.~\ref{fig:karate}(a), a widely studied social network benchmark. During the course of the study, a disagreement developed between the administrator and the club's instructor, resulting in the division of the club into two factions, represented by circles and  squares which are taken as ground truth communities for this data set.

Figure~\ref{fig:karate}(b) shows the spectra of the Laplacian and the Replicator operators.
Each spectrum contains the eigenvalues of the operator, ranked in descending order, with the largest eigenvalue in the first position. The time taken for an interaction model to reach the steady state depends on the smallest positive eigenvalue of the operator. Note that the smallest positive eigenvalue of $R$ is larger than that of $L$, implying that the non-conservative interaction model reaches steady state faster than the conservative interaction model.
We observe this empirically in Figure~\ref{fig:karate}(c) and (d) , which show the \emph{synchronization matrices} of the network at $t=1000$ under the two interaction models. Each point in the synchronization matrix represents the similarity of pairs of nodes, with red squares corresponding to higher similarity values and blue to lower. Clearly, nodes are more synchronized in the non-conservative model.

\begin{figure}[htp]
\centering
  \includegraphics[width=0.6\linewidth]{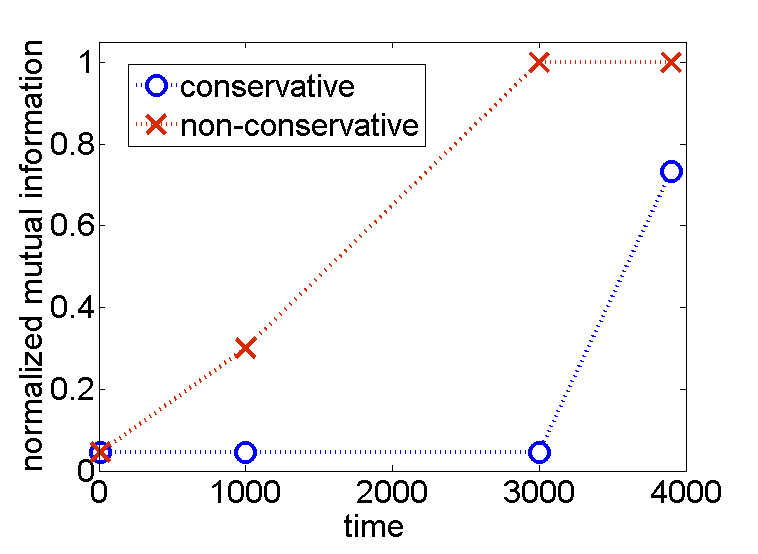}
  \caption{Evolution of the discovered community structure of the karate club network, as measured by normalized mutual information, in the conservative and non-conservative interaction models.}\label{fig:karate_nmi}
\end{figure}

We used average link hierarchical clustering algorithm to hierarchically cluster network at different times using synchronization metric as the measure of similarity. Since the ground truth communities of this network are known, we adopt normalized mutual information $MI$ as the metric for evaluating the quality of discovered communities~\cite{Danon05}. This metric measures the amount of information about actual communities that is given by the clusters found by the algorithm. When $MI=1$, the clusters are the actual communities in the network; while for $MI=0$, they are independent of the actual communities.
\remove{
Suppose our method finds a community division $X$, when the actual community division of the network is $Y$. The probability that a node is assigned to group $x$ when it actually belongs to group $y$ is $P(x,y)=N_{xy}/n$, where $N_{xy}$ is the number of nodes that were assigned to $x$ that belong to group $y$, and $n$ is the total number of nodes. The normalized mutual information is
$$
 MI(X,Y)=\frac{2I(X,Y)}{H(X)+H(Y)}, \nonumber
 $$
\noindent where standard mutual information and entropy are respectively defined as $I(X,Y)=\sum_{x,y}{P(X,Y) \log{\frac{P(X,Y)}{P(X)P(Y)}}}$, \\
$H(X)=\sum_x{P(X)\log{P(X)}}$, and  $H(Y)=\sum_y{P(Y)\log(PY)}$. When $MI=1$, the discovered communities are the actual groups in the network; while for $MI=0$, they are independent of the actual groups.
}
Figure~\ref{fig:karate_nmi} reports $MI$ scores of communities discovered at different times by the two interaction models. The non-conservative model identifies communities faster than the conservative model, and the discovered communities are purer. Conservative model assigns nodes 10 and 15 to a different community than one to which they actually belong.
\remove{
}
The non-conservative model also reveals a rich structure with a hierarchy of sub-communities. Nodes that are deeper within the hierarchy are more tightly connected, while nodes higher up, such as node $9$, $3$, $14$ and $20$, are the bridging nodes connected to both communities.

In both conservative and non-conservative models, community membership of nodes does not change much beyond $t=3899$. However, the similarity of nodes increases until the clustering procedure results in a trivial configuration, with every node equally similar to every other node. At this stage every node is assigned to the same community.

\subsection{Digg Mutual Follower Network}
{Digg} (http://digg.com) is a social news aggregator with over 3 million registered users.  Users submit links to news stories and recommend them to other users by voting on, or {digging}, them. Of the tens of thousands of daily submissions, Digg picks about a hundred to feature on its popular front page. Digg also allows users to follow other users to see the new stories they have recently submitted or voted for. We extracted data about all users who voted on stories that have been promoted to Digg's front page in June 2009, which includes users followed by these voters.\footnote{http://www.isi.edu/$\sim$lerman/downloads/digg2009.html} From this data, we reconstructed undirected mutual follower network, in which an edge between $A$ and $B$ means that user $A$ follows user $B$ \textbf{and} $B$ follows $A$.

This data set comprises of around 40K nodes and more than 360K edges.
There are 4,811 disconnected components, with the largest component comprising of $70\%$ of the nodes (27K nodes) and $96 \%$ of the edges (352K edges). The second largest component has 22 nodes. Since the inherent richness of structure of this network is largely captured by the giant component, we study this component in detail.
\remove{
\begin{figure*}[tbph]
\begin{tabular}{ccc}
\includegraphics[width=0.32\textwidth]{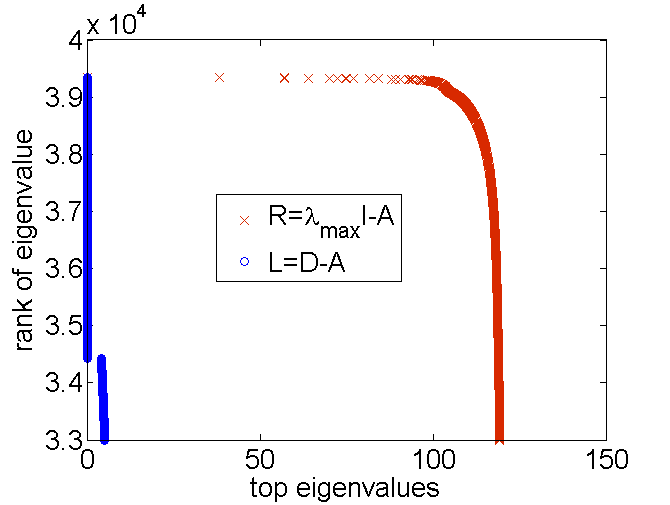}  &
\includegraphics[width=0.32\textwidth]{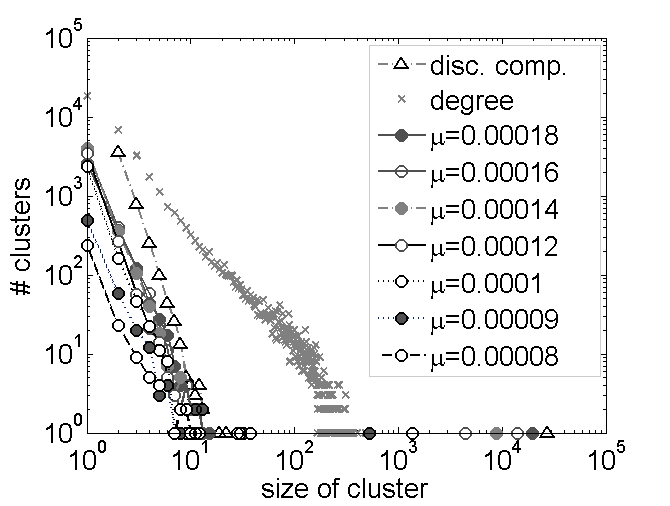} &
\includegraphics[height=0.25\textwidth]{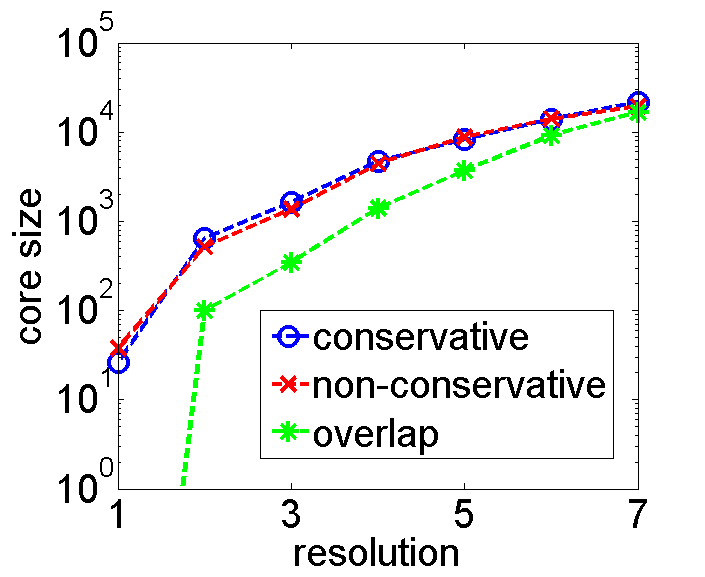} \\
(a) & (b) &(c)
\end{tabular}
\caption{ (a) Top 6000 eigenvalues of the Replicator and Laplacian operators of the Digg friendship network. (b) The long-tailed distribution (using logarithmic binning) of the components comprising the core for different similarity thresholds $\mu$ for the non-conservative model. 
(c) Comparison of core sizes found by the two interaction models at different resolution scales (specified by the similarity threshold), and the number of nodes they have in common (overlap).
}
 \label{fig:digg_stat}
\end{figure*}
}

\begin{figure}[tbh]
\centering
\begin{tabular}{c}
\includegraphics[width=0.32\textwidth]{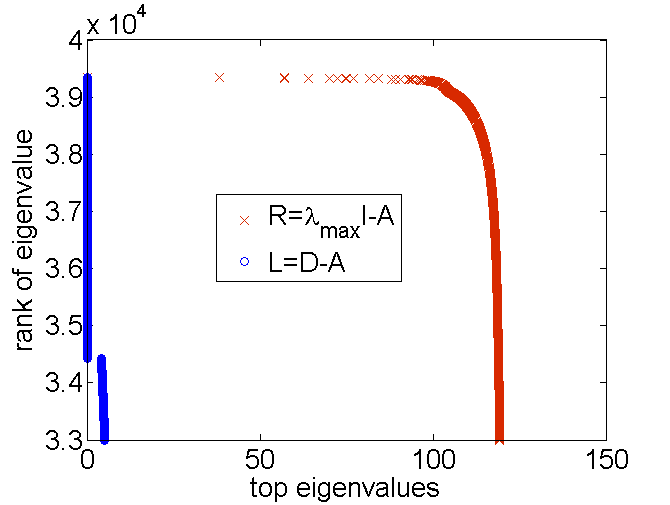} \\
(a) \\
\includegraphics[width=0.32\textwidth]{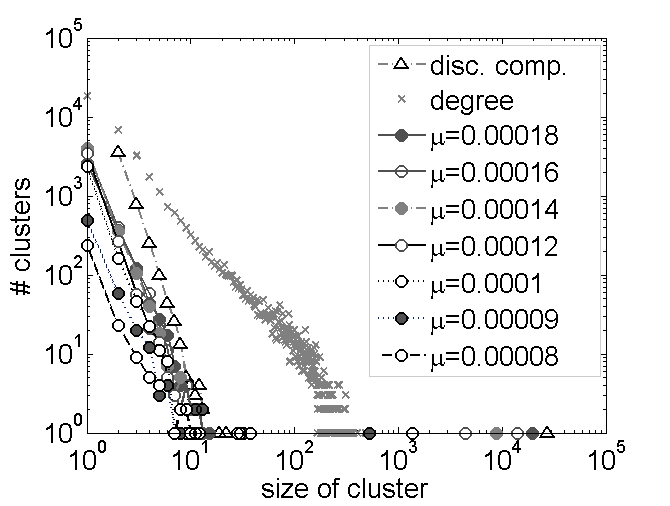} \\
(b)
\end{tabular}
\caption{(a) Top 6000 eigenvalues of the Replicator and Laplacian operators of the Digg friendship network.
(b) The long-tailed distribution (using logarithmic binning) of the components comprising the core for different similarity thresholds $\mu$ for the non-conservative model.
}
 \label{fig:digg_stat}
\end{figure}

Using the Jacobi-Davidson Algorithm
for calculating eigenvalues of a graph,  we compute more than 6K of the smallest eigenvalues of the Replicator and  Laplacian  operators and rank them in descending order (Fig.~\ref{fig:digg_stat}(a)). The two spectra a dramatically different. The smallest positive eigenvalue of $L$ is much smaller than that of $R$. This indicates that the non-conservative interaction model reaches the steady state much faster than the conservative model.

\remove
{
 \begin{figure*}[t]
\begin{tabular}{ccc}
\multicolumn{2}{c}{\includegraphics[height=0.5\textwidth]{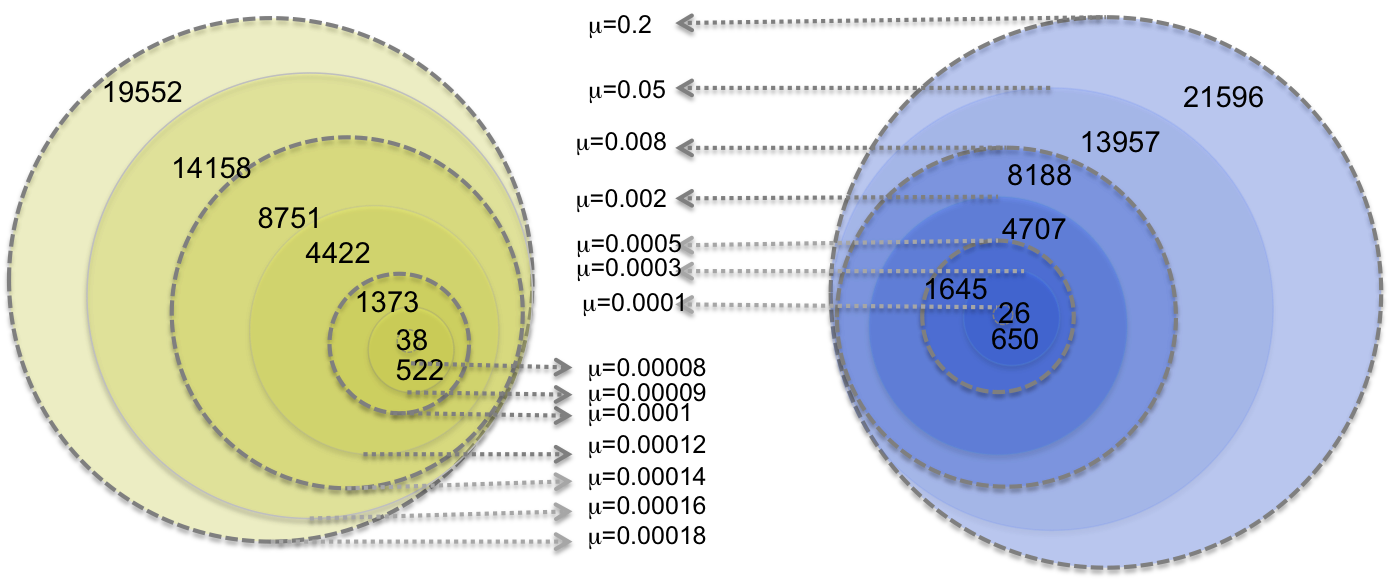} }&
\includegraphics[height=0.3\textwidth]{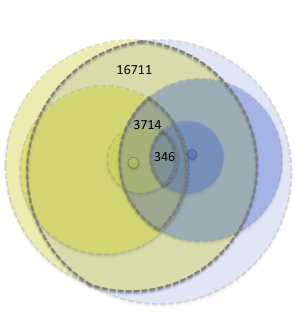} \\
(a) non-conservative & (b)conservative &(c) intersection
\end{tabular}
\caption{Pictorial representation of the fraction of nodes comprising the core at different levels of hierarchy (given by threshold $\mu$) for (a) conservative and (b) non-conservative interaction models.  The area of the circle is proportional to the number of nodes in the component. Irrespective of the nature of interaction, the community structure resembles an onion. Peeling each level of hierarchy gives a giant community (core) (shown by the circles in (a) and (b)) and a number of small communities  with a long-tailed size distribution (shown in Figure~\protect\ref{fig:digg_stat} (b) and (c)).
 (c) Overlap of selected components of comparable size (marked by grey dash line in (a) and (b)) shows that communities share few common members. }
 \label{fig:diggcore}
\end{figure*}
}

 \begin{figure}[t]
\centering
\begin{tabular}{@{}c@{}c@{}}
\includegraphics[width=0.5\linewidth]{fig/digg/overlap} &
\includegraphics[width=0.5\linewidth]{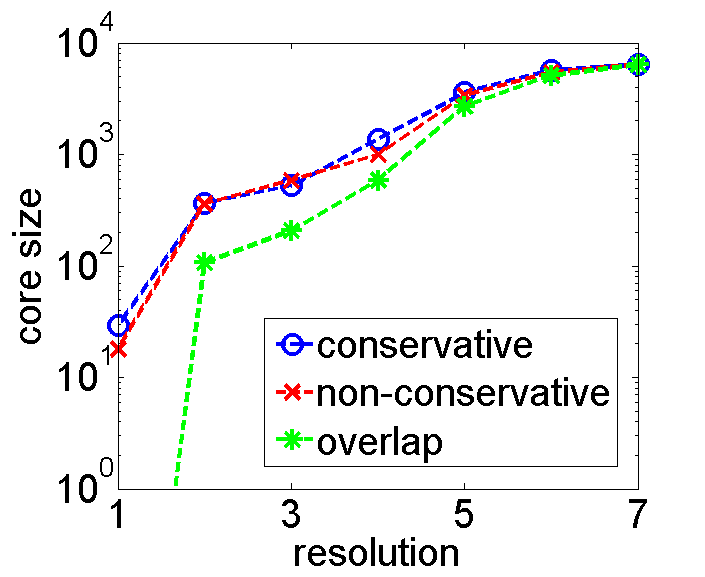} \\
(a) Digg  &
(b) Facebook
\end{tabular}
\caption{Number of nodes nodes comprising the core at different levels of hierarchy (resolution scales) found by the interaction models in the  Digg and  Facebook networks. The resolution scales correspond to similarity thresholds that give cores of comparable size. The green line shows the number of nodes that the cores have in common at that resolution scale.
}
 \label{fig:diggcore}
\end{figure}

\subsubsection{Multi-scale Structure of Digg}
We use Algorithm~\ref{alg:2} to cluster nodes at different resolutions specified by the similarity threshold $\mu$. While the overall structure changes over time, we find an intricate multi-scale organization of the network in both interaction models.
At every resolution, we find a `core and whiskers' organization~\cite{Leskovec08www}, with one giant community (core) and many small communities (whiskers). The core itself has a well-defined  structure: as we tighten the similarity threshold $\mu$, the core fragments into another large core and many small communities  with a long-tailed size distribution. This process continues until the core fragments into some number of small communities. 

The community structure of Digg, therefore, resembles an onion, with multiple layers of whiskers.  This paradigm is captured in Figure~\ref{fig:diggcore}(a), which shows core sizes at different resolution scales at time $t=100$.
At later times, at any given resolution the core grows until $t=t_{eq}$, when it forms a giant component for every resolution scale. However, the composition of the core remains almost time invariant, i.e. the core at a coarser resolution at time $t_1$ is very similar to a core at some  finer resolution at later time $t_2$.
We chose the threshold parameters $\mu$ that give comparable size cores at each resolution scale for the interaction models.

Using the non-conservative interaction model, all thresholds above $\mu=0.0004$ produce a single component with about 27K nodes.  At a finer resolution (smaller $\mu$), the number of communities increases. As illustrated in Figure~\ref{fig:diggcore}(a), at $\mu=0.00018$,  $76\%$ of these nodes form a giant component or the core.  In addition, there are several small communities, whose sizes have a long-tailed distribution (Fig.~\ref{fig:digg_stat}(b)). At $\mu=0.00016$, the core again divides into one large community, with  $72\%$ of the nodes, and many small communities, whose sizes also have a long-tailed distribution, as shown in Figure \ref{fig:digg_stat}(b). Increasing the resolution scale further to $\mu=0.00014$, we discover that the core found at  $\mu=0.00016$ breaks down once more into one giant component comprising of $62\%$ of the nodes, and so on.
A similar organization is discovered using the conservative model and though at later times.

While the onion-like organization discovered by both interaction models is similar, its composition is different. Figure~\ref{fig:diggcore}(a) shows the overlap of the membership of comparable-size cores found by the two models. For example, the size of the giant component discovered by non-conservative interaction model for $\mu=0.00018$ \remove{ ($19552$ nodes)} is comparable to the size of the core discovered by the conservative interaction model for $\mu=0.2$\remove{ ($19552$ nodes)}; however, they share only about $80\%$ of the nodes.
Core overlap decreases to about $40\%$ at $\mu=0.00014$ for non-conservative interaction model ($\mu=0.008$ for conservative model), and keeps on decreasing as we fine-tune the resolution scale. Finally, the largest component at $\mu=0.00008$ for non-conservative and $\mu=0.0001$ for conservative models (resolution scale 1) do not have any nodes in common.

\subsubsection{Empirical Evaluation}
While the two interaction models discover different structures in the Digg network, in the absence of ground truth communities for this network, it is challenging to say which model is correct. However, user activity provides an independent source of evidence for evaluating the quality of communities. We use this evidence to gain more insight into the structure of the Digg network, and show that the non-conservative model is better suited for studying it.

\begin{figure}[tbh]
\begin{tabular}{@{}c@{}c@{}}
  \includegraphics[width=0.5\linewidth]{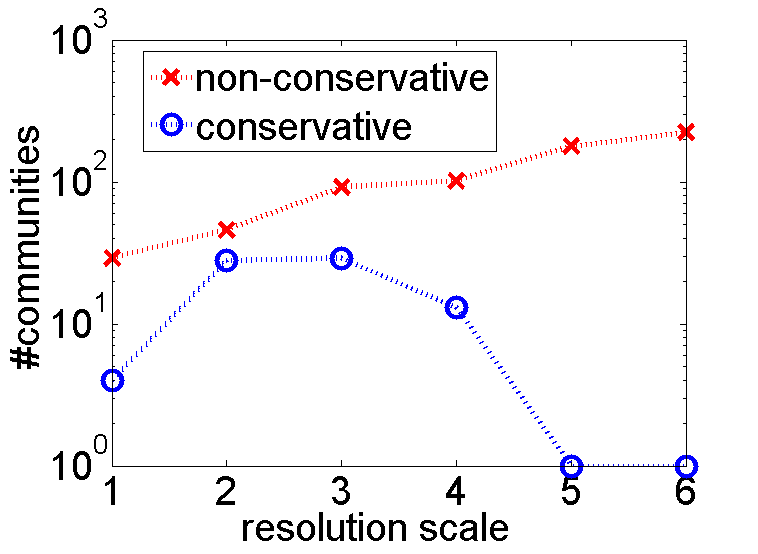} &
  \includegraphics[width=0.5\linewidth]{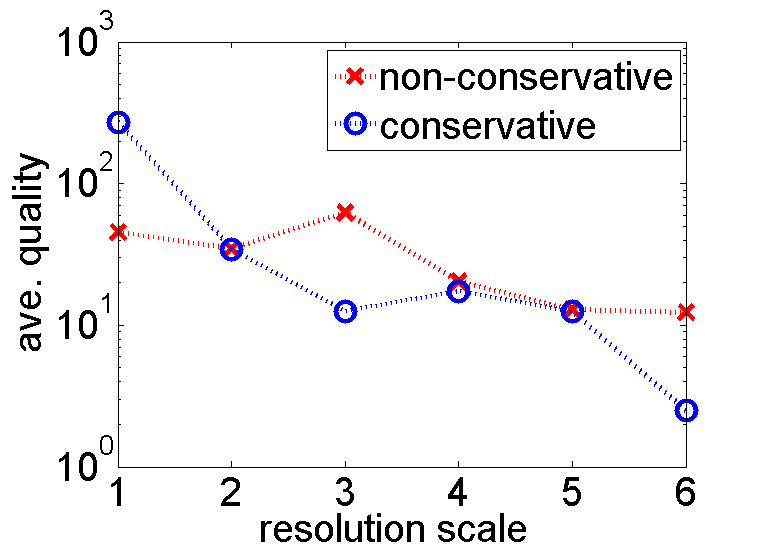}  \\
  (a) & (b)
\end{tabular}
  \caption{Evaluation of communities found in the Digg mutual follower graph at $t=100$ by the two interaction models. (a) Number of small communities found at different resolutions specified by the similarity threshold parameter. The smallest resolution corresponds to smallest value of the similarity threshold. (b) Average quality of communities at each scale, as measured by the number of co-votes.}\label{fig:digg_results}
\end{figure}

We propose an empirical measure of  community quality based on user activity. Members of the same community are likely to share the same information, interests, and attributes~\cite{Granovetter73}. As a consequence, they are likely to behave  in a similar manner, which on Digg translates into voting for the same news stories. We measure similarity of two Digg users by the number of  stories for which they both voted, i.e., \emph{co-votes}. Then, averaging over co-votes of all pairs of community members, we obtain a number that quantifies the quality of the community. We focus on small components (whiskers) of at least size three isolated from the core at different resolutions. Non-conservative interaction model assigned 3,712 users to such small communities. In contrast, the conservative interaction model assigned just 449 users to small communities. The rest of the users fragmented into isolated pairs or singletons.

Figure~\ref{fig:digg_results}(a) shows the number of small communities resolved by the two interaction models at different scales.  Figure~\ref{fig:digg_results}(b) reports the average community quality at each resolution scale, as measured by the number co-votes between pairs of community members.  Community quality increases at finer resolution scales, producing tighter communities in the center of the `onion' as expected. Members of the innermost communities (resolution scale 1), are much more similar than members of the outer communities (resolution scales 5, 6). Except for these innermost communities, the average quality of communities found by the non-conservative model is better than that found by the conservative model.  The difference at resolution scale 1 is driven by the two outliers in the conservative model. The first of these is a community of 26 users, with more than 300 co-votes on average, and the other is a community of nine with more than 600 co-votes. In addition to co-voting on an extraordinary number of stories (600 is nearly 20\% of all stories in our data set), these users are also highly interlinked. The first group forms a 13-core (a cluster in which each node is linked to at least 13 other nodes), and the second group forms a 4-core. These users also share many friends. While we cannot say whether these groups represent the often-rumored voting blocs on Digg, their activity does appear to be anomalous.  One way such activity could arise is if each member of the group navigated to the profiles of other group members and voted for the stories that appeared on that profile, e.g., the stories that member submitted or voted for. Such browsing can be represented by one-to-one interactions; therefore, conservative model is best at finding it. Non-conservative models describe information diffusion through broadcasts of recent votes to followers, and finds communities arising from this information sharing behavior. To summarize, non-conservative model finds many more small communities of higher quality than the conservative model, though the latter seems to pick out some anomalous groups of users.

\begin{figure}[tbhp]
\begin{tabular}{@{}c@{}c@{}}
  \includegraphics[width=0.5\linewidth]{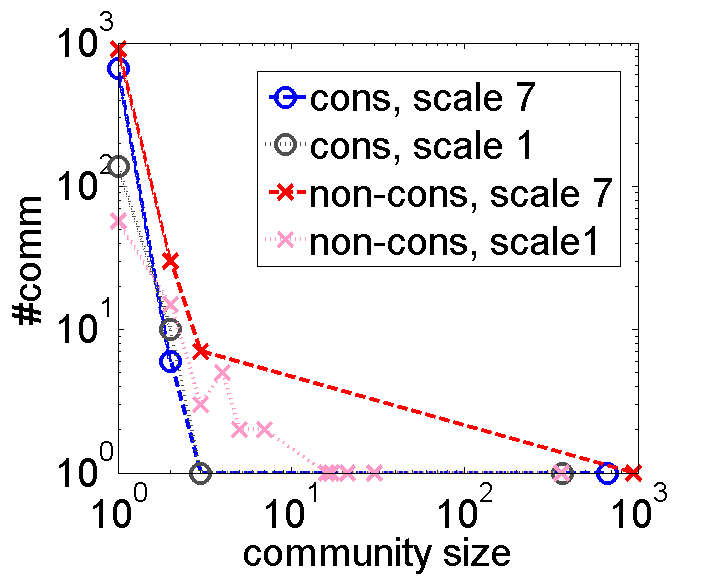} &
  \includegraphics[width=0.5\linewidth]{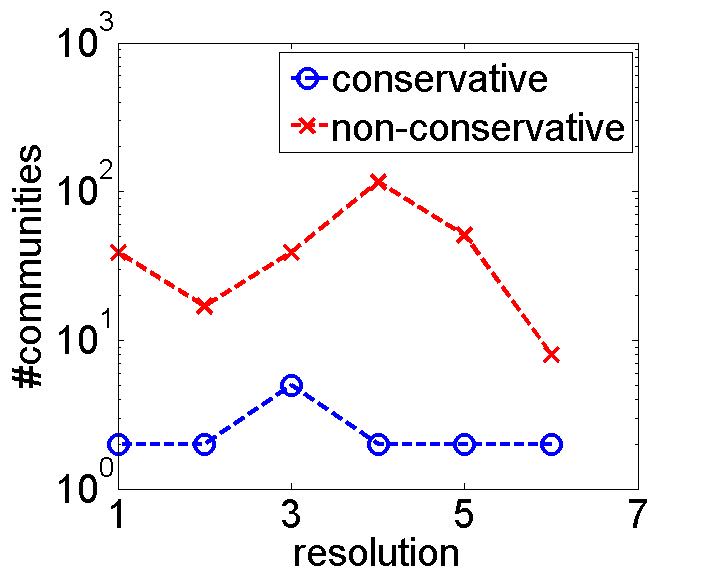}  \\
  (a) & (b)
\end{tabular}
  \caption{Distribution of communities in the Facebook network for American University at $t=100$. (a) Comparison of size distribution of small communities found by the two interaction models at the coarsest and finest resolution scales. (b)  Number of small communities at different resolutions. The smallest resolution corresponds to highest similarity between individuals.}\label{fig:fb_comms}
\end{figure}

\subsection{Facebook}
We performed our analysis on a data set containing a snapshot of the Facebook networks as of September 2005 \cite{facebook,facebooklong}. Each user in this data set has four descriptive features: status (e.g., student, faculty, staff, and so on),  major, dorm or house, and graduation year. We use these features to empirically evaluate the quality of the discovered communities, in a sense that a good community should consist of individual who are similar according to these features.
While this data set contains more than 100 colleges and universities, we present here the analysis of the network for American University, which comprises of 6,386 nodes and more than 200K edges.

\subsubsection{Multi-scale Structure of Facebook}
We use Algorithm~\ref{alg:2} to cluster nodes at different resolution scales specified by the similarity threshold $\mu$. As with Digg, we find an onion-like, multi-scale organization in the structures discovered by conservative and non-conservative interaction models for the Facebook networks
underpinning the generality of the observed structure.
At each resolution, we discover a giant community (core) and many small communities (whiskers) with a long tailed size distribution (Fig.~\ref{fig:fb_comms}(a)). Just as on Digg, there is little overlap in membership between cores found by the two interaction models at finer resolutions (Fig.~\ref{fig:diggcore}(b)).

As on Digg, many nodes participate in small, clique-like communities. However, while 1,320 nodes contribute to the formation of such communities in the non-conservative interaction model, only 32 nodes participate in such communities in the conservative interaction model. The remaining users are fragmented into isolated pairs or singletons. As in the Digg data set, non-conservative model found many more communities than the conservative model.  Figure~\ref{fig:fb_comms}(b) shows the number of communities discovered at each resolution scale for conservative and non-conservative models.

\subsubsection{Empirical Evaluation}

\begin{figure}[tbhp]
\begin{tabular}{@{}c@{}c@{}}
    \includegraphics[width=0.5\linewidth]{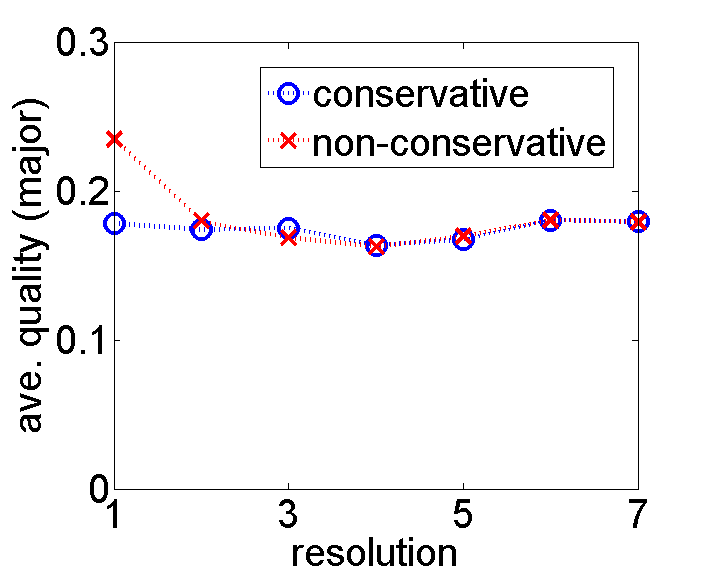} &
  \includegraphics[width=0.5\linewidth]{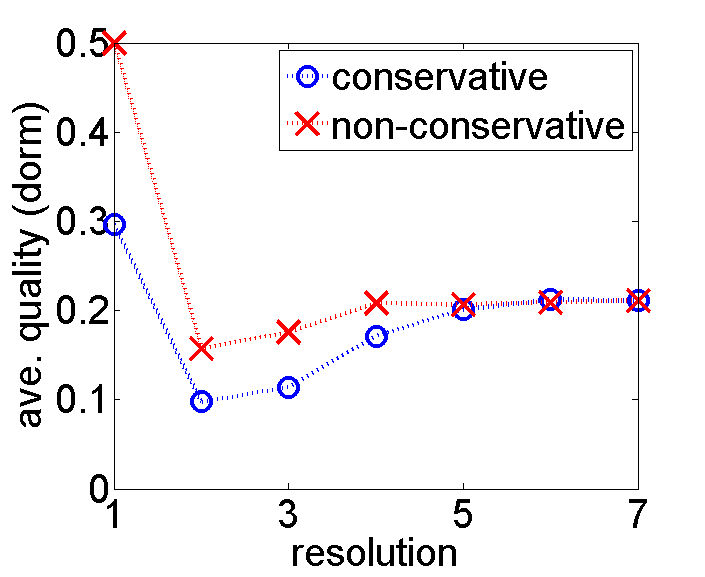}  \\
  major &  dorm\\
      \includegraphics[width=0.5\linewidth]{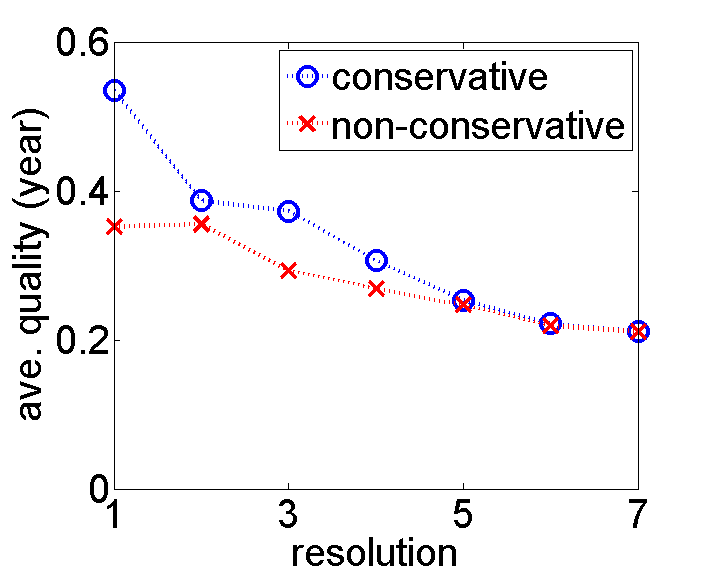} &
  \includegraphics[width=0.5\linewidth]{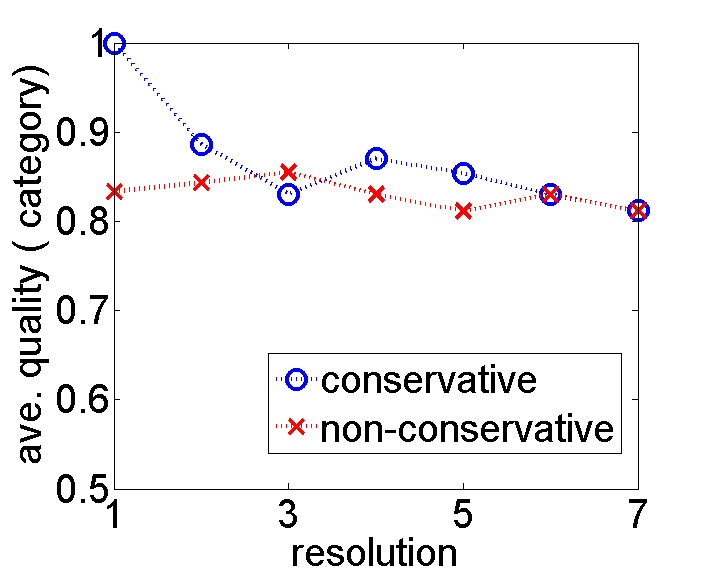}  \\
  year & category
\end{tabular}
  \caption{Evaluation of communities found in the Facebook network of American University at $t=100$ by the two interaction models.  Average quality of communities at each resolution scale, as measured by the probability of occurrence of the most frequent value of features major, dorm, year, category of individual. }\label{fig:fb_results}
\end{figure}

We measure quality of the community discovered at different resolution scales using the four features enumerated above namely: major, dorm, year and category of individual.
We measure the prevalence of the most popular value of some feature among community members. If the community is pure, the quality will be high.
For example, the quality of a community with respect to the dorm feature gives the largest fraction of community members that belong to the same dorm.
Figure~\ref{fig:fb_results} reports quality of communities found by the two models at different resolution scales with respect to those features.
We find that, overall, the quality increases as we tighten the similarity threshold (decrease the resolution scale), irrespective of the feature under consideration. However, the characteristics of the community structure discovered by conservative and non-conservative interaction models vary significantly.
At finer resolution scales, non-conservative model finds communities of individuals who are more likely to have the same major and belong to the same dorm. Conservative models, on the other hand, are more likely to put into the same community individuals who belong to the student category and are in the same year. Though the type of interactions may differ from college to college, it is reasonable to assume that students who belong to the same year will have more face to face (conservative) interactions, while students who have the same major or live in a dorm, may meet in study groups, or organized events, increasing chances for one-to-many (non-conservative) interaction.

In summary, regardless of the interaction process, we observe a roughly scale invariant organization in the real-world social networks. At  almost every resolution scale, we find a large component and many small components with a long-tailed size distribution.  Thus, Digg and Facebook's structure resembles an onion. Peeling each layer reveals another, almost self-similar structure with a core and many smaller communities. However, the composition of communities depends on the interaction process, and is different for the conservative and non-conservative interaction models.

\section{Related Work}
Community detection is an extremely active research area, with a variety of methods proposed, including spectral clustering, graph partitioning and modularity maximization~\cite{Fortunato2010Community}. We show in this paper that these methods  can be expressed in terms of the generalized linear interaction model that assumes a specific type of interaction. We also demonstrate that to get the full picture of network's emergent structure, community detection method must account for the dynamic process occurring on the network.

 It might be argued that taking interactions into account eventually leads to a weighted graph and off-the shelf community detection algorithms for weighted graphs~\cite{Fortunato2010Community} might be applied.
However, like in the unweighted case, application of a community detection method on a weighted graph without taking the nature of interaction process into account, might lead to unsatisfactory results.
For example, if conductance minimization algorithm is applied on a weighted graph, whose weights are a consequence of a non-conservative process, the structure detected might differ significantly from ground truth. Our method on the other hand learns the weights from the interaction process and then detects structure dynamically. Learning the underlying interaction process from the activity logs of nodes of the network and using this process to determine the community structure is the course of future work.


Several community detection methods implicitly takes dynamic interactions into account. These include spin models, random walk models and synchronization. Spin models \cite{spin} imply that the interaction is ferromagnetic i.e. it favors spin alignment. As we show in this paper, random walk and Kuramoto synchronization models~\cite{Kuramoto} are both conservative in nature, with the former expressed in terms of the normalized Laplacian, and the latter in terms of the graph Laplacian.
Arenas et al.~\cite{Arenas06} studied the relationship between topological and community structure of complex networks using the Kuramoto model of synchronization. They created a threshold graph at some point in time where an edge exists between nodes only if their similarity exceeds some threshold. They defined communities as disconnected components of the threshold graph.  We, on the other hand, explore different types of interactions and show how these reveal different hierarchical community structures in real-world complex networks. We also introduce a process-independent similarity metric.
Hu et al.~\cite{Hu} found communities based on signaling interactions. They described the interactions by an operator $\mathcal{L}(A)=(I+A)$ and used K-means clustering and F-statistics to find the optimal clusters at a some point of time. However, it can be shown mathematically that the process they defined will never reach a steady state. Our non-conservative interaction model treats signaling interactions in a principled way.

Community detection methods are used to reveal the structure of complex networks. Leskovec et al.~\cite{Leskovec08www} found `core and whiskers' structure of real-world networks using conductance-based methods and argued that these methods cannot reveal any further structure in the giant core. Song~\cite{Song} claimed that there exist self-repeating patterns in complex networks at all length scales.  Our results corroborate this claim, as we show a repeating `core and whiskers' pattern in the Digg social network at many different length scales.

It can be shown that  some of the interaction models described above not only solve certain regularized Semi-Definite Programs but also give fast solutions to these problems\cite{perry2011}.

\section{Conclusion}
Our work highlights the importance of dynamic interactions in the analysis of network structure and provides a framework for unifying some of the existing community detection methods. We argue that in order to understand network structure,  not only its topology but also the nature of interactions between nodes should be taken into consideration. We have proposed a novel non-conservative interaction model inspired by distributed synchronization of a network of coupled oscillators. We also presented a new  formulation of similarity which we used in multi-scale analysis of network structure and an activity-based metric to measure the quality of communities in a real-world network. Our decentralized approach to the community detection is fast and scalable.

Our study of the community structure of real-world social networks revealed a complex `onion'-like organization. Peeling each level of hierarchy gives a core and many small components, regardless of the interaction model. However, different interactions lead to different views of this multi-scale organization.

In future, we would like to investigate the effect of non-zero $\omega$ on the interaction models.
Also, we would like to investigate the  ergodicity of interaction models and the spectral properties of the different operators.
Our work offers a framework for understanding the role of dynamic processes in the measurement of network structure. We hope that our investigations inspire others to explore the relationships between network structure, topology and dynamics.

\bibliographystyle{abbrv}
\small

\bibliography{rumig-synchronization,rumig-www2012,krisl-refs}  
\end{document}